\documentclass{article}
\usepackage{amsmath}
\usepackage{tikz}
\usepackage[linesnumbered, ruled]{algorithm2e}
\usepackage{authblk}
\usepackage[margin=2cm, includefoot, footskip=30pt]{geometry}

\title{Memory depth of finite state machine strategies for the iterated prisoner's dilemma}
\author[1]{T.J. Gaffney*}
\author[1]{Marc Harper}
\author[2]{Vincent A. Knight}
\affil[1]{Google Inc., Mountain View, CA, United States of America}
\affil[2]{Cardiff University, School of Mathematics, Cardiff, United Kingdom}

\begin{document}

\maketitle

\begin{abstract}
  We develop an efficient algorithm to determine the memory-depth of finite state machines and apply the algorithm to
  a collection of iterated prisoner's dilemma strategies. The calculation agrees with the memory-depth of other
  representations of common strategies such as Tit-For-Tat, Tit-For-2-Tats, etc. which are typically represented by
  lookup tables.
  Our algorithm allows the complexity of finite state machine based strategies to be characterized on the same
  footing as memory-\(n\) strategies.
\end{abstract}

\section{Introduction}
The Prisoner's dilemma is a classic game-theoretic model of competitive interactions.
As a one shot-game it has a non-cooperative Nash equilibrium, however with repeated play cooperative strategies can be effective.
The Iterated Prisoner's Dilemma (IPD) has been studied extensively in tournaments, population games, and via systematic searches for
effective strategies to the repeated game.
Though strategies can be arbitrarily complex, Robert Axelrod and collaborators showed that simple strategies
such as Tit-for-Tat (TFT) could defeat many more-complicated strategies \cite{axelrod1981evolution} in tournament
settings.

Subsequent research has investigated the effect of complexity on the performance of strategies \cite{Ashlock2013}.
In many cases memory-depth, defined by the amount of rounds of history used to decide the next action, has served as a proxy for complexity \cite{Ashlock2013}.
A strategy requiring $n$ rounds of the history is called a memory-\(n\) strategy. TFT is a memory-one strategy depending
only on the opponent's previous action, as are many recently defined strategies such as zero-determinant strategies \cite{Press2012}.
It has been shown, in some contexts, that longer-memory strategies are more evolutionary stable \cite{JiaweiLi2014}.
More generally, strategies with long memories have been observed to outperform shorter-memory strategies \cite{adami2013evolutionary, Knight2018, lee2015art} in various contexts.

Researchers have also evolved or otherwise trained strategies to compete in various IPD contexts \cite{Ashlock2006b, Harper2017}.
Such processes can produce strategies of substantially varying complexity.
Generally these processes require an encoding of a subset of the space of all strategies into a representation that can breed and mutate.
Two such representations are lookup tables and finite state machine (FSM) \cite{Ashlock2013}. 
Lookup tables have a fixed memory-depth and one can easily restrict a population to individuals of 
fixed depth with an evolutionary algorithm.
\cite{Leas2016} showed that longer such strategies (up to four states) performed better in a pool with shorter strategies.

In contrast, the transitions of a finite state machine may vary with simple variations, and the memory-depth is not
simply the number of states, though some authors have used the number of states as a measure of complexity.
The literature has not previously defined memory-depth of FSM strategies, perhaps due to the non-triviality of the calculation.
Ashlock et al \cite{Ashlock2006b} showed that fewer-state strategies tend to appear in evolved populations.
Nevertheless, the memory-depth of a finite state machine is often infinite, requiring the entirety of the history of play.

Since strategies may be represented or implemented in many ways, a representation agnostic definition of
complexity such as memory-depth is preferrable to a parameter-based definition such as the number of states
of a finite state machine. This paper defines an efficient algorithm to calculate the memory-depth of a FSM strategy,
allowing a comparability of complexity to strategies represented in other ways.  

\section{Iterated Prisoner's Dilemma, Memory Depth, and Finite State Machines}

The Prisoner's Dilemma is a simple game where each player chooses to cooperate or defect.
The two players choose their move at the same time, and receive the cooresponding payoff shown in Table \ref{PD}.
The IPD has two players play each other repeatedly.   Players form strategies to decide when to cooperate and
when to defect, taking into account what happened in previous turns.

\begin{table}
\caption{Payoffs Matrix for Prisoner's Dilemma}
\label{PD}
\[
\begin{array}{ccccc}
 & \vline & \text{Cooperate} & \vline & \text{Defect} \\
\hline
\text{Cooperate} & \vline & (3,3) & \vline & (0,5) \\
\hline
\text{Defect} & \vline & (5,0) & \vline & (1,1) \\
\end{array}
\]
\end{table}

A \textit{strategy} is any set of rules for how to play the prisoner's dilemma; for example a simple (but effective) strategy, called Tit-For-Tat (TFT),
 acts by copying the opponent's previous move.  A strategy may use any 
general rule to respond to past turns.  For example, the Go-By-Majority strategy is a strategy that cooperates if
the opponent has cooperated at least half the time.  The \textit{memory-depth} of a strategy is the smallest amount
of recent history that a strategy needs to know in order to choose the next move.  For example, TFT only needs to know
what the opponent did last turn to make a decision; a variant Tit-For-Two-Tats (TF2T) defects only after two opponent
defects, and therefore has a memory-depth of two.
Generally speaking, we say a strategy has memory-depth $n$ if $n$ is the smallest number for which the strategy would
work the same if it only had knowledge of the previous $n$ moves (including its own previous $n$ moves).
Some strategies need their entire history; we say that these strategies have infinite memory-depth.
For example, the Go-By-Majority strategy needs to know the entire play history at any point in order to know what to
do next.  We say that such a strategy has infinite memory-depth.

Some strategies are most-easily described by how they respond to a few previous moves.  One example
is the Fortress-3 strategy, shown in Figure \ref{FortressLookup}.
(When a strategy is defined with a table of previous $n$ moves, like this, we call the table a \textit{lookup representation}.)
This strategy defects unless the previous
three opponent moves are defects.  The memory depth is therefore $3$.\footnote{This strategy only uses the opponent's
previous three moves, but in general a memory-depth $3$ strategy could also use information about its own previous three
moves.}
With a lookup representation, it is easy to see what the memory-depth is, but not all strategies are represented as lookups.
The definition of memory-depth is independent of strategy representation.  In this paper we explore memory-depth of
strategies represented as an FSM (defined below).

\begin{table}[!htbp]
\caption{Lookup table Fortress-3}

\label{FortressLookup}
\[
\begin{array}{cc}
\underline{\textbf{Previous Three Opponent Moves}} & \underline{\textbf{Strategy's Next Move}} \\
C, C, C                                      & D                                   \\
C, C, D                                      & D                                   \\
C, D, C                                      & D                                   \\
C, D, D                                      & D                                   \\
D, C, C                                      & D                                   \\
D, C, D                                      & D                                   \\
D, D, C                                      & D                                   \\
D, D, D                                      & C                                   \\
\end{array}
\]
\end{table}

An FSM is specified by $(S, \Sigma, \Gamma, s_0, \delta, \omega)$, where $S$ is the set of
states, $\Sigma$ is a set of
input symbols, $\Gamma$ is a set of output symbols, $s_0$ is an initial state, $\delta: S\times\Sigma\to S$ is a
state-transition function, and $\omega:S\times\Sigma\to\Gamma$ is an output function.
It starts in state $s_0$, and (when in state $s$) responds to an input, $\sigma\in\Sigma$, by outputting
$\omega(s, \sigma)$, and transitioning to state $\delta(s, \sigma)$ \footnote{Notation from Soucha \cite{Soucha}}.
We further say that two states, $s_i$ and $s_j$, have the same \textit{response sets} if
$\omega(s_i, \sigma)=\omega(s_j, \sigma), \text{ for all }\sigma\in\Sigma$, that is, if the mapping from input action
to output action is the same for the two states.

\begin{figure}
\centering
\caption{Finite state machine for Fortress-3}
\vspace*{5mm}
\label{FsmFortress}
\begin{tikzpicture}[scale=0.15]
\tikzstyle{every node}+=[inner sep=0pt]
\draw [black] (24.3,-12) circle (3);
\draw (24.3,-12) node {$1$};
\draw [black] (38.4,-12) circle (3);
\draw (38.4,-12) node {$2$};
\draw [black] (38.4,-25) circle (3);
\draw (38.4,-25) node {$3$};
\draw [black] (24.3,-25) circle (3);
\draw (24.3,-25) node {$4$};
\draw [black] (21.531,-10.876) arc (275.63354:-12.36646:2.25);
\draw (17.58,-6.47) node [above] {$D/D$};
\fill [black] (23.51,-9.12) -- (23.93,-8.27) -- (22.93,-8.37);
\draw [black] (38.4,-15) -- (38.4,-22);
\fill [black] (38.4,-22) -- (38.9,-21.2) -- (37.9,-21.2);
\draw (38.9,-18.5) node [right] {$C/D$};
\draw [black] (35.763,-13.415) arc (-68.81005:-111.18995:12.21);
\fill [black] (35.76,-13.42) -- (34.84,-13.24) -- (35.2,-14.17);
\draw (31.35,-14.74) node [below] {$C/D$};
\draw [black] (26.815,-10.383) arc (114.7926:65.2074:10.814);
\fill [black] (26.82,-10.38) -- (27.75,-10.5) -- (27.33,-9.59);
\draw (31.35,-8.89) node [above] {$D/D$};
\draw [black] (36.19,-22.97) -- (26.51,-14.03);
\fill [black] (26.51,-14.03) -- (26.75,-14.94) -- (27.43,-14.21);
\draw (29.11,-18.99) node [below] {$D/D$};
\draw [black] (35.4,-25) -- (27.3,-25);
\fill [black] (27.3,-25) -- (28.1,-25.5) -- (28.1,-24.5);
\draw (31.35,-25.5) node [below] {$C/C$};
\draw [black] (23.544,-27.891) arc (13.08562:-274.91438:2.25);
\draw (17.76,-30.59) node [below] {$C/C$};
\fill [black] (21.55,-26.16) -- (20.65,-25.85) -- (20.88,-26.83);
\draw [black] (24.3,-22) -- (24.3,-15);
\fill [black] (24.3,-15) -- (23.8,-15.8) -- (24.8,-15.8);
\draw (23.8,-18.5) node [left] {$D/D$};
\end{tikzpicture}
\end{figure}
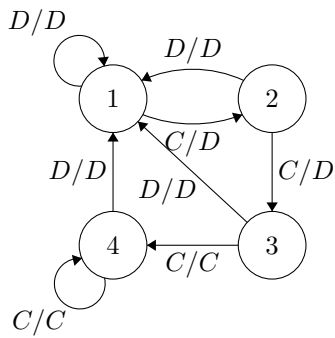

Some strategies are naturally described as a FSM.  For example, the Fortress-3 strategy described above
can be described by the FSM in Figure \ref{FsmFortress}.
State 1 represents a state where the last opponent move was a defect, state 2 represents a state where the last
opponent move was a cooperation, state 3 reperesents a state where the last two opponent moves were cooperations,
and state 4 represents a state where three or more of the opponent's most resent moves are
cooperations.\footnote{This could also be presented as a three-state FSM, since states 3 and 4 behave the same.}
Using this as an example, we can see that each state has two outgoing edges, one for each possible opponent action:
In state 3 the outgoing edges are C/C going to state 4 and D/D going to state 1.
This means that in state 3, the strategy responds to a cooperation by cooperating and switching to state 4, and to a
defection with a defection and a switch to state 1.
Similarly, in state 1 the outgoing edges are C/D going to state 2 and D/D
returning to state 1. The strategy responds to a cooperation by defecting and
switching to state 4, and to a defection by defecting and staying in state 1.

There is not is an obvious way to read memory-depth from a strategy represented as an FSM.
Generally there is no close connection between number of states and
memory-depth; the number of states is dependent on the representation, while memory-depth exists for any strategy
independent of representation.  Memory-depth is the amount of history required to decide the next move.  For FSMs the next
move is given immediately by the current state, which may suggest that the next move can be determined immmediately.
However the current state is an internal variable, based on the history.
Just as GoByMajority updates an internal variable tracking
portion of time opponent cooperated, a FSM updates an internal variable tracking state.\footnote{
Generally determining the current state is sufficient but not necessary to decide the next move, because multiple states
may have the same result set; in that case we don't need to determine which of those states the strategy is in.
The existing literature sometimes requires ``minimal" machines, which precludes such pairs of states, but this is too
restrictive for our purposes.
}  And like with
GoByMajority's internal state, it may take all the history to reconstruct an FSM's internal state (as
demonstrated in \ref{FsmInfinite}, below).  This paper describes a way to calculate the
memory-depth for any strategy that can be represented by a FSM.

Though our algorithm is general, we apply it here to strategies for the IPD.  The FSMs define the
strategies: the inputs are the opponent's previous action and the outputs are the strategy's reaction.
The only possible moves are Cooperate (C) and Defect (D) so $\Sigma=\Gamma=\{C, D\}$.
A machine may have any finite number of states.

\section{Modified PDS Algorithm}

There is a body of research that explores the following question:
what is the shortest sequence of input actions for which the output actions can distinguish which state the machine is in?
Such a sequence is called \emph{distinguishing sequence}.
There have been many answers to this question, which are reviewed in Soucha \cite{Soucha}.
Our memory-depth question differs in a few ways:  Firstly our memory question looks backwards instead of forwards.
Secondly we only want to be able to distinguish between states with different response sets, not to distinguish all states.
Thirdly we seek not the shortest distinguishing sequence, rather we must find a depth for which all sequences of that
depth are distinguishing.
This is equivalent to finding the longest \textit{non-distinguishing sequence}, and adding one.

Despite these differences, one of the approaches in the existing literature can be easily adapted, though inefficiently.
The Preset Distinguishing Sequence (PDS) algorithm~\cite{Soucha} \cite{Deshmukh1994} builds all sequences of possible
future moves, stopping when it finds the shortest distinguishing sequence.
As a modified approach, we could instead build backwards, calling a sequence distinguishing when all the terminal states
have the same response set, continuing until the longest non-distinguishing sequence has been found.

The algorithm involves building a \textit{reverse distinguishing tree}, which is a predecessor tree.
Formally, the algorithm has the following steps:

\begin{enumerate}
\item Make a root node for a tree, containing a \textit{path} for each state $s\in S$. A path is just an object with a \textit{terminal} and a \textit{current} state, written as $\left(s_t, s_c\right)$.  We initially set both terminal and current states to $s$.
\item For each leaf in the tree (initially just the root), if all the terminal states in the leaf have the same response set, then we mark that leaf as "resolved."
\item For each non-resolved leaf, walk backward from current states, while keeping track of terminal states.  That is, for each non-resolved leaf, $P$, build child nodes, $\left\{C_\sigma\right\}$ in the following way:  For each path, $\left(s_t, s_c\right)$ in $P$, and each transition $\delta(s_{c'}, \sigma)=s_c$, put a path $\left(s_t, s_c\right)$.
\item Repeat steps 2 and 3 until all leaves are resolved.  The height of the tree (in the number of nodes) is the memory-depth.
\end{enumerate}

\begin{figure}
\centering
\caption{Reverse distinguishing tree for Fortress-3}
\vspace*{5mm}
\label{RdtFortress}
\begin{tikzpicture}
\tikzstyle{level 1}=[sibling distance=30mm]
\tikzstyle{level 2}=[sibling distance=20mm]
\node[rectangle, rounded corners, draw](state){$1,2,3,4$}
  child {node[rectangle, rounded corners, draw](state1){$4$}
    edge from parent node[above, sloped]{\scriptsize C/C}
  }
  child {node[rectangle, rounded corners, draw](state22){$2,3$}
    child{node[](state31){}}
    child{node[rectangle, rounded corners, draw](state32){$3$} edge from parent node[left]{\scriptsize C/D}}
    child{node[](state33){}}
    child{node[rectangle, rounded corners, draw](state34){$2$} edge from parent node[above, sloped]{\scriptsize D/D}}
    edge from parent node[left]{\scriptsize C/D}
  }
  child {node[](state23){$$}
    edge from parent node[right]{\scriptsize D/C}
  }
  child {node[rectangle, rounded corners, draw](state24){$1$}
    edge from parent node[above, sloped]{\scriptsize D/D}
  };
\end{tikzpicture}
\end{figure}
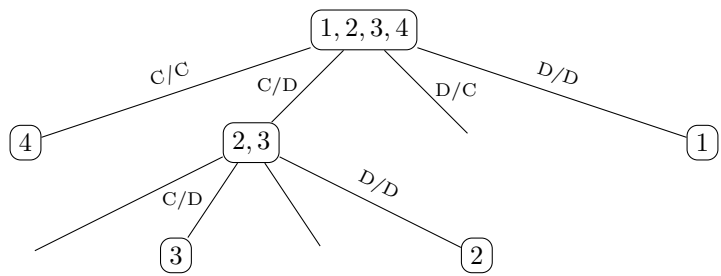

To see how the algorithm works with an example, we will continue with our Fortress-3 strategy from above:
The only state that can follow C/C is 4.
The only state that can follow D/D is 1.
However a C/D could indicate either a transition from 1 to 2 or from 2 to 3.
At this point, state 1 (terminal state 2) can only have D/D feed into it, and state 2 (terminal state 3) can only have
C/D feed into it.
We build the reverse distinguishing tree in Figure \ref{RdtFortress}; in drawing the graph, we only write the terminal
states for the paths of a leaf.

\begin{figure}
\centering
\caption{Shifted reverse distinguishing tree for Fortress-3}
\vspace*{5mm}
\label{SrdfFortress}
\begin{tikzpicture}
\tikzstyle{level 1}=[sibling distance=50mm]
\tikzstyle{level 2}=[sibling distance=45mm]
\tikzstyle{level 3}=[sibling distance=40mm]
\node[rectangle, rounded corners, draw](state){$1,2,3,4$}
  child {node[rectangle, rounded corners, draw](state1){$4$}
    edge from parent node[above, sloped]{\scriptsize /C, */}
  }
  child {node[rectangle, rounded corners, draw](state22){$1,2,3$}
    child{node[rectangle, rounded corners, draw](state32){$2,3$}
		child{node[rectangle, rounded corners, draw](state42){$3$} edge from parent node[above, sloped]{\scriptsize /C, D/}}
		child{node[rectangle, rounded corners, draw](state44){$2$} edge from parent node[above, sloped]{\scriptsize /D, D/}}
		edge from parent node[above, sloped]{\scriptsize /D, C/}}
    child{node[rectangle, rounded corners, draw](state32){$1$} edge from parent node[above, sloped]{\scriptsize /D, D/}}
    edge from parent node[above, sloped]{\scriptsize /D, */}
  };
\end{tikzpicture}
\end{figure}
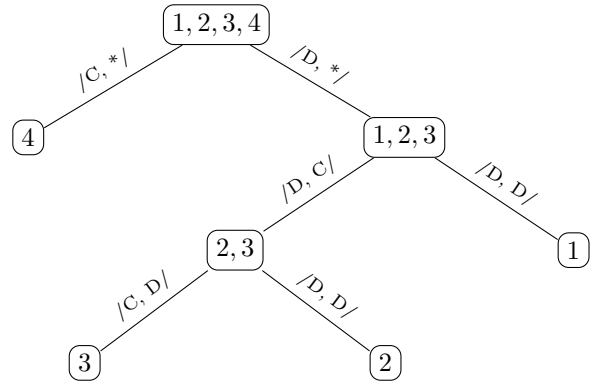

When all the leaves are resolved (as in Figure \ref{RdtFortress}) the algorithm halts.
We can then determine the memory-depth.
For this example, at the bottom-most leaf, in order to distinguish between response sets, we need the input actions.
This is the opponent's action from three turns ago.
Therefore the memory-depth is three.
Generally, the memory-depth cannot be inferred by the height of the tree alone, because the output action is the
strategy's own action two turns ago.
Hence it matters how the bottom layer is distinguished.
To handle this we shift the nodes so that an edge is on the same level of memory as shown in Figure \ref{SrdfFortress}.

With this setup it will always be the case that the height of the fully resolved tree (or rather height minus one since
we're counting edges) is the memory-depth of the strategy.
(The * is used to specify that opponent action in the most recent turn won't affect the state the strategy is on.)

\begin{figure}
\centering
\caption{Memit reverse distinguishing tree for Fortress-3}
\vspace*{5mm}
\label{MrdtFortress}
\begin{tikzpicture}
\tikzstyle{level 0}=[sibling distance=30mm]
\tikzstyle{level 1}=[sibling distance=30mm]
\tikzstyle{level 2}=[sibling distance=30mm]
\node[](par){}
child {node[rectangle, rounded corners, draw](state01){${}_D1_*,{}_D2_*,{}_D3_*$}
  child {node[rectangle, rounded corners, draw](state12){${}_D1_D,{}_D2_D,{}_D3_D,{}_D4_D$}}
  child {node[rectangle, rounded corners, draw](state11){${}_D1_C,{}_D2_C$}
    child {node[rectangle, rounded corners, draw](state22){${}_D1_D,{}_D2_D,{}_D3_D,{}_D4_D$}}
    child {node[rectangle, rounded corners, draw](state21){${}_D1_C$}}
  }
}
child {node[rectangle, rounded corners, draw](state00){${}_C4_*$}};
\end{tikzpicture}
\end{figure}
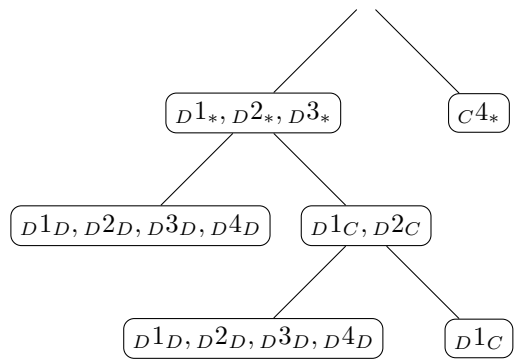

We will find it convenient later to rewrite this so that actions are matched with the state, as we do in Figure \ref{MrdtFortress}.
We call a state along with the incoming/outgoing actions a \textit{memit} (or memory unit).
With this representation we write current states instead of terminal states.
It's less clear from the representation that the nodes with ${}_D1_D,{}_D2_D,{}_D3_D,{}_D4_D$ are resolved, but they are
in this case because they all represent paths with the same terminal node.
A useful property of this representation is that the same children follow from a fixed parent, regardless of the position in the graph.
This repeated calculation is the motivation for the more efficient algorithm of the next section.

\begin{figure}
\centering
\caption{FSM graph of infinite strategy}
\vspace*{5mm}
\label{FsmInfinite}
\begin{tikzpicture}[scale=0.15]
\tikzstyle{every node}+=[inner sep=0pt]
\draw [black] (19.4,-17) circle (3);
\draw (19.4,-17) node {$1$};
\draw [black] (30.3,-17) circle (3);
\draw (30.3,-17) node {$2$};
\draw [black] (16.72,-18.323) arc (324:36:2.25);
\draw (12.15,-17) node [left] {$D/C$};
\fill [black] (16.72,-15.68) -- (16.37,-14.8) -- (15.78,-15.61);
\draw [black] (32.98,-15.677) arc (144:-144:2.25);
\draw (37.55,-17) node [right] {$D/D$};
\fill [black] (32.98,-18.32) -- (33.33,-19.2) -- (33.92,-18.39);
\draw [black] (27.8,-18.625) arc (-67.92828:-112.07172:7.85);
\fill [black] (27.8,-18.62) -- (26.87,-18.46) -- (27.25,-19.39);
\draw (24.85,-19.7) node [below] {$C/C$};
\draw [black] (21.897,-15.37) arc (112.15544:67.84456:7.831);
\fill [black] (21.9,-15.37) -- (22.83,-15.53) -- (22.45,-14.61);
\draw (24.85,-14.29) node [above] {$C/C$};
\end{tikzpicture}
\end{figure}
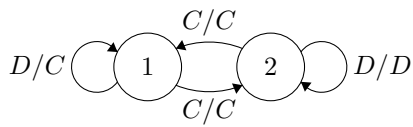

Since the memory-depth of FSMs are often infinite, there is no guarantee of convergence.  We look at a simple example:
Consider a strategy (shown in Figure \ref{FsmInfinite}) that in state 1 will cooperate (regardless of opponent action),
and in state 2 will respond in kind to either a cooperate or a defect, and with every opponent cooperation the strategy
switches states.
An arbitrarily long chain of C/C could occur without us knowing the state, and therefore we wouldn't know how to respond to a defect.
If we trace our algorithm, we would find a never-ending branch of C/C and so this algorithm doesn't have a good way to
detect infinite-memory strategies.  However in the next section, we see that finite-memory strategies will converge after at most $S^2$ steps.

The Modified PDS algorithm, though intuitive, is slow as the number of branches can grow exponentially.
Moreover since we have to run the algorithm many times before we can conclude that a strategy has infinite memory,
it is impractical. The following algorithm addresses both these points.

\section{Memit Pair Algorithm}

A better approach is to look at pairs of memits.
If there is a sequence is that is not distinguishable, then there will be a sequence of memit pairs that will be
indistinguishable at every step.
So we look instead for such a sequence of pairs.
To do this, we build a graph where the nodes are all memit pairs that are not distinguishable by their actions.
An edge between a pair of memits $(A, B)$ to another pair of memits $(C, D)$ exists if there's a transition from $A$ to
$C$ and a from $B$ to $D$ (or $A\to D$, $B\to C$.)
Then we find the longest path in this graph.

The algorithm works as follows.

\begin{enumerate}
\item Make a \textit{memit graph}, where the nodes are all possible memits, represented as triples: (incoming player action, state, outgoing opponent action).  Make an edge from memit $(a, X, b)$ to $(c, Y, d)$ if the response to an opponent action $b$ when in state $X$ is to respond $c$ and move to state $Y$.  Said mathematically, $\omega(X, b)=c$, $\delta(X, b)=Y$.
\item Make a \textit{memit pair graph}, where the nodes are all pair of memits $(a, X, b)$ and $(c, Y, d)$ for which the states are different ($X\neq Y$), but the actions are the same ($a=c, b=d$).  Label this $(a, X, Y, b)$ and count this the same as $(a, Y, X, b)$, making only one node for both.  Make an edge from memit-pair $(a, X, Y, b)$ to memit-pair $(c, Z, W, d)$, if there is are edges (in the memit graph) $(a, X, b)\to(c, Z, d)$ and $(a, Y, b)\to(c, W, d)$.  [Or if there are edges $(a, X, b)\to(c, W, d)$ and $(a, Y, b)\to(c, Z, d)$.]
\item For each node in the memit pair graph for which the states in the memit pair have distinct response sets, find the longest path ending on that node.  If there is a path ending at that memit pair which has a cycle, then count as infinite.
\item The number of nodes in the longest of these paths plus one is the memory-depth.
\end{enumerate}

\begin{figure}
\centering
\caption{Memit graph of Fortress-3 strategy}
\vspace*{5mm}
\label{MemitFortress}
\begin{tikzpicture}[scale=0.2]
\definecolor{red}{rgb}{1,0.3,0.3}
\definecolor{green}{rgb}{0.3,1.0,0.3}
\definecolor{blue}{rgb}{0.45,0.45,1.0}
\definecolor{gray}{rgb}{0.7,0.7,0.7}
\tikzstyle{every node}+=[inner sep=0pt]
\draw [black, fill=red] (8.7,-25.4) circle (3);
\draw (8.7,-25.4) node {${}_D1_D$};
\draw [black, fill=green] (8.7,-34.1) circle (3);
\draw (8.7,-34.1) node {${}_D1_C$};
\draw [black, fill=blue] (18.6,-21.2) circle (3);
\draw (18.6,-21.2) node {${}_C4_D$};
\draw [black, fill=gray] (28.4,-21.2) circle (3);
\draw (28.4,-21.2) node {${}_C4_C$};
\draw [black, fill=red] (18.6,-30.2) circle (3);
\draw (18.6,-30.2) node {${}_D3_D$};
\draw [black, fill=green] (28.4,-30.2) circle (3);
\draw (28.4,-30.2) node {${}_D3_C$};
\draw [black, fill=red] (18.6,-39) circle (3);
\draw (18.6,-39) node {${}_D2_D$};
\draw [black, fill=green] (28.4,-39) circle (3);
\draw (28.4,-39) node {${}_D2_C$};
\draw [black] (15.84,-22.37) -- (11.46,-24.23);
\fill [black] (11.46,-24.23) -- (12.39,-24.38) -- (12,-23.46);
\draw [black] (16.77,-23.58) -- (10.53,-31.72);
\fill [black] (10.53,-31.72) -- (11.41,-31.39) -- (10.62,-30.78);
\draw [black] (15.9,-28.89) -- (11.4,-26.71);
\fill [black] (11.4,-26.71) -- (11.9,-27.51) -- (12.34,-26.61);
\draw [black] (15.81,-31.3) -- (11.49,-33);
\fill [black] (11.49,-33) -- (12.42,-33.17) -- (12.05,-32.24);
\draw [black] (16.83,-36.57) -- (10.47,-27.83);
\fill [black] (10.47,-27.83) -- (10.53,-28.77) -- (11.34,-28.18);
\draw [black] (15.91,-37.67) -- (11.39,-35.43);
\fill [black] (11.39,-35.43) -- (11.88,-36.23) -- (12.33,-35.34);
\draw [black] (5.781,-24.759) arc (285.34019:-2.65981:2.25);
\fill [black] (7.43,-22.69) -- (7.7,-21.79) -- (6.74,-22.05);
\draw [black] (15.684,-39.608) arc (-90.14098:-142.52519:7.205);
\fill [black] (15.68,-39.61) -- (14.89,-39.11) -- (14.88,-40.11);
\draw [black] (26.728,-41.478) arc (-42.29614:-165.63944:10.367);
\fill [black] (26.73,-41.48) -- (25.82,-41.73) -- (26.56,-42.41);
\draw [black] (26.17,-37) -- (20.83,-32.2);
\fill [black] (20.83,-32.2) -- (21.09,-33.11) -- (21.76,-32.37);
\draw [black] (28.4,-36) -- (28.4,-33.2);
\fill [black] (28.4,-33.2) -- (27.9,-34) -- (28.9,-34);
\draw [black] (26.19,-28.17) -- (20.81,-23.23);
\fill [black] (20.81,-23.23) -- (21.06,-24.14) -- (21.74,-23.4);
\draw [black] (28.4,-27.2) -- (28.4,-24.2);
\fill [black] (28.4,-24.2) -- (27.9,-25) -- (28.9,-25);
\draw [black] (29.1,-18.295) arc (194.19443:-93.80557:2.25);
\fill [black] (31.13,-19.99) -- (32.03,-20.28) -- (31.78,-19.31);
\draw [black] (25.4,-21.2) -- (21.6,-21.2);
\fill [black] (21.6,-21.2) -- (22.4,-21.7) -- (22.4,-20.7);
\end{tikzpicture}
\end{figure}
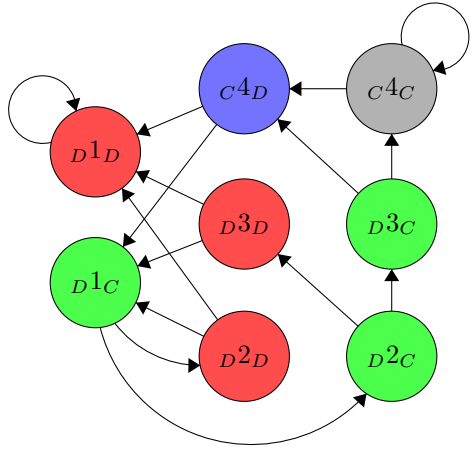

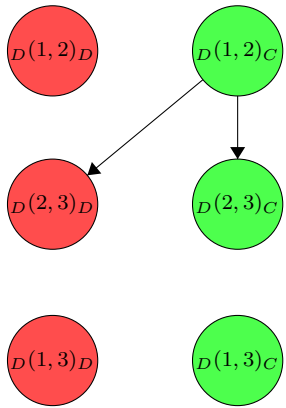
\begin{figure}
\centering
\caption{Memit-pair graph of Fortress-3 strategy}
\vspace*{5mm}
\label{MemitPairFortress}
\begin{tikzpicture}[scale=0.2]
\definecolor{red}{rgb}{1,0.3,0.3}
\definecolor{green}{rgb}{0.3,1.0,0.3}
\tikzstyle{every node}+=[inner sep=0pt]
\draw [black, fill=red] (21.1,-14.1) circle (3);
\draw (21.1,-14.1) node {\footnotesize ${}_D(1, 2)_D$};
\draw [black, fill=green] (33.4,-14.1) circle (3);
\draw (33.4,-14.1) node {\footnotesize ${}_D(1, 2)_C$};
\draw [black, fill=red] (21.1,-24.3) circle (3);
\draw (21.1,-24.3) node {\footnotesize ${}_D(2, 3)_D$};
\draw [black, fill=green] (33.4,-24.3) circle (3);
\draw (33.4,-24.3) node {\footnotesize ${}_D(2, 3)_C$};
\draw [black, fill=red] (21.1,-34.7) circle (3);
\draw (21.1,-34.7) node {\footnotesize ${}_D(1, 3)_D$};
\draw [black, fill=green] (33.4,-34.7) circle (3);
\draw (33.4,-34.7) node {\footnotesize ${}_D(1, 3)_C$};
\draw [black] (33.4,-17.1) -- (33.4,-21.3);
\fill [black] (33.4,-21.3) -- (33.9,-20.5) -- (32.9,-20.5);
\draw [black] (31.09,-16.02) -- (23.41,-22.38);
\fill [black] (23.41,-22.38) -- (24.34,-22.26) -- (23.71,-21.49);
\end{tikzpicture}
\end{figure}

As an example, consider a run of the algorithm on the Fortress-3 strategy from above.
We first draw the full memit graph in Figure \ref{MemitFortress}, and the graph of memit pairs in Figure \ref{MemitPairFortress}.
We see that there is a path, for example, from the memit pair $(D, 1, 2, C)$ to $(D, 2, 3, D)$ because in the memit
graph $(D, 1, C)$ maps to $(D, 2, D)$ and $(D, 2, C)$ maps to $(D, 3, D)$.

\begin{figure}
\centering
\caption{Memit-pair graph of infinte memory strategy}
\vspace*{5mm}
\label{MemitPairInfinite}
\begin{tikzpicture}[scale=0.2]
\definecolor{blue}{rgb}{0.45,0.45,1.0}
\definecolor{gray}{rgb}{0.7,0.7,0.7}
\tikzstyle{every node}+=[inner sep=0pt]
\draw [black, fill=gray] (32.6,-18.9) circle (3);
\draw (32.6,-18.9) node {\footnotesize ${}_C(1,2)_C$};
\draw [black, fill=blue] (32.6,-30) circle (3);
\draw (32.6,-30) node {\footnotesize ${}_C(1,2)_D$};
\draw [black] (32.6,-21.9) -- (32.6,-27);
\fill [black] (32.6,-27) -- (33.1,-26.2) -- (32.1,-26.2);
\draw [black] (33.812,-16.169) arc (183.80557:-104.19443:2.25);
\fill [black] (35.51,-18.2) -- (36.34,-18.65) -- (36.27,-17.65);
\end{tikzpicture}
\end{figure}
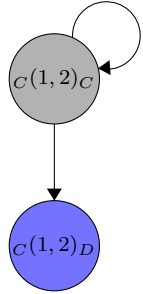

To see how this would work for the infinite-memory case, we will repeat this calculation for the two-state strategy
of Figure~\ref{FsmInfinite}.
The memit graph is shown in Figure \ref{MemitInfinite} and the memit-pair graph is shown in Figure \ref{MemitPairInfinite}.
What makes this memory infinite is the loop between ${}_C1_C$ and ${}_C2_C$, which forms a self-loop on this memit pair.

\begin{figure}
\centering
\caption{Memit graph of infinte memory strategy}
\vspace*{5mm}
\label{MemitInfinite}
\begin{tikzpicture}[scale=0.2]
\definecolor{red}{rgb}{1,0.3,0.3}
\definecolor{green}{rgb}{0.3,1.0,0.3}
\definecolor{blue}{rgb}{0.45,0.45,1.0}
\definecolor{gray}{rgb}{0.7,0.7,0.7}
\tikzstyle{every node}+=[inner sep=0pt]
\draw [black, fill=gray] (27,-15.6) circle (3);
\draw (27,-15.6) node {${}_C1_C$};
\draw [black, fill=blue] (37.5,-15.6) circle (3);
\draw (37.5,-15.6) node {${}_C1_D$};
\draw [black, fill=gray] (20.5,-24.3) circle (3);
\draw (20.5,-24.3) node {${}_C2_C$};
\draw [black, fill=red] (44.1,-24.3) circle (3);
\draw (44.1,-24.3) node {${}_D2_D$};
\draw [black, fill=blue] (27,-33.6) circle (3);
\draw (27,-33.6) node {${}_C2_D$};
\draw [black, fill=green] (37.5,-33.6) circle (3);
\draw (37.5,-33.6) node {${}_D2_C$};
\draw [black] (19.685,-21.446) arc (-178.66963:-254.85913:5.886);
\fill [black] (19.69,-21.45) -- (20.17,-20.63) -- (19.17,-20.66);
\draw [black] (22.3,-21.9) -- (25.2,-18);
\fill [black] (25.2,-18) -- (24.33,-18.34) -- (25.13,-18.94);
\draw [black] (27,-18.6) -- (27,-30.6);
\fill [black] (27,-30.6) -- (27.5,-29.8) -- (26.5,-29.8);
\draw [black] (38.018,-12.657) arc (197.74616:-90.25384:2.25);
\fill [black] (40.15,-14.22) -- (41.07,-14.45) -- (40.76,-13.5);
\draw [black] (34.5,-15.6) -- (30,-15.6);
\fill [black] (30,-15.6) -- (30.8,-16.1) -- (30.8,-15.1);
\draw [black] (23.17,-22.93) -- (34.83,-16.97);
\fill [black] (34.83,-16.97) -- (33.89,-16.89) -- (34.35,-17.78);
\draw [black] (30,-33.6) -- (34.5,-33.6);
\fill [black] (34.5,-33.6) -- (33.7,-33.1) -- (33.7,-34.1);
\draw [black] (29.64,-32.17) -- (41.46,-25.73);
\fill [black] (41.46,-25.73) -- (40.52,-25.68) -- (41,-26.55);
\draw [black] (35.99,-31.01) -- (28.51,-18.19);
\fill [black] (28.51,-18.19) -- (28.48,-19.13) -- (29.35,-18.63);
\draw [black] (37.5,-30.6) -- (37.5,-18.6);
\fill [black] (37.5,-18.6) -- (37,-19.4) -- (38,-19.4);
\draw [black] (46.78,-22.977) arc (144:-144:2.25);
\fill [black] (46.78,-25.62) -- (47.13,-26.5) -- (47.72,-25.69);
\draw [black] (42.36,-26.75) -- (39.24,-31.15);
\fill [black] (39.24,-31.15) -- (40.11,-30.79) -- (39.29,-30.21);
\end{tikzpicture}
\end{figure}
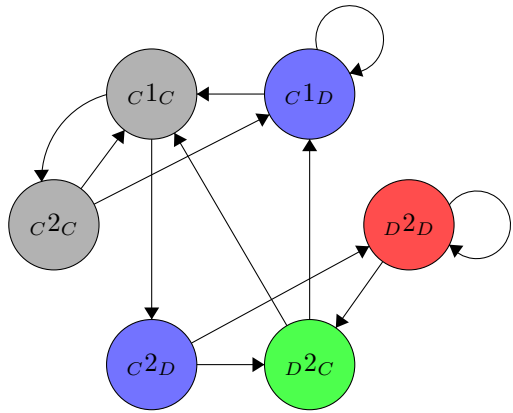

\begin{figure}
\centering
\caption{Memit-pair graph of two-state TFT}
\vspace*{5mm}
\label{MemitPairTft}
\begin{tikzpicture}[scale=0.2]
\tikzstyle{every node}+=[inner sep=0pt]
\draw [black] (21.4,-34.9) circle (3);
\draw (21.4,-34.9) node {${}_C(1,2)_C$};
\draw [black] (33.1,-34.9) circle (3);
\draw (33.1,-34.9) node {${}_C(1,2)_D$};
\draw [black] (21.4,-44.6) circle (3);
\draw (21.4,-44.6) node {${}_D(1,2)_C$};
\draw [black] (33.1,-44.6) circle (3);
\draw (33.1,-44.6) node {${}_D(1,2)_D$};
\draw [black] (24.4,-34.9) -- (30.1,-34.9);
\fill [black] (30.1,-34.9) -- (29.3,-34.4) -- (29.3,-35.4);
\draw [black] (18.489,-34.227) arc (284.71059:-3.28941:2.25);
\fill [black] (20.16,-32.18) -- (20.44,-31.28) -- (19.48,-31.53);
\draw [black] (31.447,-37.401) arc (-37.50007:-63.17841:21.288);
\fill [black] (24.16,-43.44) -- (25.1,-43.52) -- (24.65,-42.63);
\draw [black] (33.1,-37.9) -- (33.1,-41.6);
\fill [black] (33.1,-41.6) -- (33.6,-40.8) -- (32.6,-40.8);
\draw [black] (21.4,-41.6) -- (21.4,-37.9);
\fill [black] (21.4,-37.9) -- (20.9,-38.7) -- (21.9,-38.7);
\draw [black] (23.112,-42.139) arc (141.43954:117.88198:23.074);
\fill [black] (30.36,-36.13) -- (29.42,-36.06) -- (29.89,-36.94);
\draw [black] (30.1,-44.6) -- (24.4,-44.6);
\fill [black] (24.4,-44.6) -- (25.2,-45.1) -- (25.2,-44.1);
\draw [black] (36.045,-45.107) arc (107.97263:-180.02737:2.25);
\fill [black] (34.49,-47.25) -- (34.26,-48.16) -- (35.21,-47.85);
\end{tikzpicture}
\end{figure}
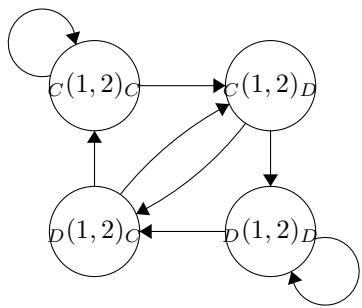

It is important for this example that state 1 and state 2 have different response sets, which is required in step 3 of the algorithm.  To demonstrate why, we consider TFT written with two states, shown in Figure \ref{FsmTft}.  TFT responds in kind to the opponent's previous move; it has memory-depth one and is usually represented with just one state.  However when we write it as two states, it yields the memit-pair graph in Figure \ref{MemitPairTft}, which has loops, which seems to imply an infinite memory.  This can be explained by the fact that states 1 and 2 have the same response set, and therefore there is no path ending in a node with a pair of states having different response sets.\footnote{Because there are no such paths, this must be handled specially, as described in step \ref{step4} below.}

\begin{figure}
\centering
\caption{FSM graph of two-state TFT}
\vspace*{5mm}
\label{FsmTft}
\begin{tikzpicture}[scale=0.15]
\tikzstyle{every node}+=[inner sep=0pt]
\draw [black] (19.4,-17) circle (3);
\draw (19.4,-17) node {$1$};
\draw [black] (30.3,-17) circle (3);
\draw (30.3,-17) node {$2$};
\draw [black] (16.72,-18.323) arc (324:36:2.25);
\draw (12.15,-17) node [left] {$D/D$};
\fill [black] (16.72,-15.68) -- (16.37,-14.8) -- (15.78,-15.61);
\draw [black] (32.98,-15.677) arc (144:-144:2.25);
\draw (37.55,-17) node [right] {$C/C$};
\fill [black] (32.98,-18.32) -- (33.33,-19.2) -- (33.92,-18.39);
\draw [black] (27.8,-18.625) arc (-67.92828:-112.07172:7.85);
\fill [black] (27.8,-18.62) -- (26.87,-18.46) -- (27.25,-19.39);
\draw (24.85,-19.7) node [below] {$C/C$};
\draw [black] (21.897,-15.37) arc (112.15544:67.84456:7.831);
\fill [black] (21.9,-15.37) -- (22.83,-15.53) -- (22.45,-14.61);
\draw (24.85,-14.29) node [above] {$D/D$};
\end{tikzpicture}
\end{figure}
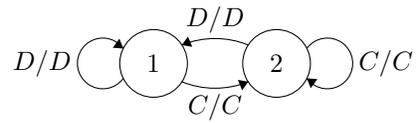

\section{Algorithm Pseudocode}

In this section, we show the implementation of the algorithm in pseudocode.
The algorithm is implemented and tested in full in the Axelrod library for Python
\cite{axelrodproject} (version 4.7.0 at the time of writing). \footnote{
In the algorithm, we formally reverse the graph, but here we just describe the algorithm as looking backwards.}

We break down the algorithm into five steps:

\begin{enumerate}
\item \label{step1} Reduce graph to reachable states from the initial state, deleting unreachable nodes, and edges to or from them.
\item \label{step2} Build memit graph, as described above.
\item \label{step3} Build memit-pair graph, as described above.
\item \label{step4} If there are no tied memits, then manually check if the memory-depth is $0$ (all actions are the same) or $1$ (otherwise).\footnote{This is a special case, since, as we see below, we add one to the length of the longest path to get the memory-depth.  So there's no way that the general algorithm can detect a memory-$0$ strategy.} 
\item \label{step5} Otherwise, for each memit-pair node, for which the memits have a different response set, find the longest path (in the memit-pair graph) ending at that node.  If there is a loop in the path, then we say that the memory-depth is infinite.  The memory-depth is the length (in nodes) of the longest such path plus one.
\end{enumerate}

Step \ref{step4} is self-explanatory, and steps \ref{step1} and \ref{step5} are well-known graph calculations, so these are omited here.  Algorithms for steps \ref{step2} and \ref{step3} are shown below.  For these algorithms, we will call memits $(in\_action, state, out\_action)$.  For algorithm \ref{Alg2}, we say that two memits are equal if their $in\_action$ and $out\_action$ are the same, and the states are different.

\begin{algorithm}
  \KwIn{FSM given as list of graph transitions, $(current\_state, opponents\_last\_action, next\_state, response)$.}
  \KwOut{MemitGraph}    
  \caption{Build memit graph, per step \ref{step2}.}
  \DontPrintSemicolon
  \BlankLine
  Initialize $incoming\_actions$, a map from states to incoming actions\;
  \ForEach{\textup{Transition in FSM}}{
  	Add $response$ action to $incoming\_actions$ with key $next\_state$\;
  }
  \BlankLine
  \ForEach{\textup{Transition in FSM}}{
    \ForEach{$in\_action$\textup{ in }$incoming\_actions[current\_state]$}{
      \ForEach{$out\_action$\textup{ in {Defect and Cooperate}}}{
        Set $start\_node$ to $(in\_action, current\_state, opponents\_last\_action)$\;
        Set $end\_node$ to $(response, next\_state, out\_action)$\;
        Add an edge from $start\_node$ to $end\_node$ to Memit Graph\;
      }
    }
  }
\end{algorithm}

\begin{algorithm}
  \KwIn{MemitGraph, given as adjacency lists, $\{successors[memit]\}$}
  \KwOut{MemitPairGraph}    
  \caption{Build memit-pair graph, per step \ref{step3}.}
  \label{Alg2}
  \DontPrintSemicolon
  \BlankLine
  \ForEach{$x\in\textup{Memits}, y\in\textup{Memits} : x=y \textup{ }$}{
    \textup{Add $(x, y)$ to the memit-pair graph}\;
    \ForEach{$sx\in successors[x], sy\in successors[y] : sx=sy \textup{ }$}{
      \textup{Add an edge from $(x, y)$ to $(sx, sy)$}\;
    }
  }
\end{algorithm}

The runtime of the first two steps in the algorithm is linear in the $S$, number of states of the FSM.  Building the memit-pair graph involves looping through all pairs of memits, which will take $O(S^2)$ time.  The resulting memit-pair graph will have $O(S^2)$ vertices, and $O(S^2)$ edges, since each vertex can only have at most two edges leaving it.  Step \ref{step5} loops through every vertex, and for each vertex finds the longest path using a search that will, at most, reach every vertex.  The total runtime of this step is $O(S^4)$, which is the longest step.

As we trace a path in Step \ref{step5} of the above algorithm, we can see the work that would have to be done in the Modified PDS algorithm of the previous section.  A path of length $n$ on the memit-pair represents a unresolved complete tree with depth $n$.\footnote{We can see now that if the Modified PDS algorithm does not converge after a depth of $S^2$, then it must be an infinite-memory strategy.}  In the worst case scenario, the modified PDS algorithm would need a tree of size $O(2^{S^2})$ to resolve a pair of states.

\section{Data Analysis}

The Axelrod library currently has 17 strategies to the IPD that are represented as FSMs, from various
literature sources and in some cases trained by evolutionary algorithms.
We list these in Table \ref{data_analysis} along with the number of states used to build the FSM and the memory-depth.
We see that the strategies naturally described by a FSM are often infinite memory-depth.

\begin{table}[!htbp]
\label{memory_lengths}
\caption{Memory-depths and number of states for Axelrod FSM strategies}
\label{data_analysis}
\[
\begin{array}{lcc}
\underline{\textbf{Strategy Name}} & \underline{\textbf{States}} & \underline{\textbf{Memory Depth}} \\
\text{Fortress 3} & 3 & 2 \\
\text{Fortress 4} & 4 & 3 \\
\text{Predator} & 9 & \infty \\
\text{Pun1} & 2 & \infty \\
\text{Raider} & 4 & \infty \\
\text{Ripoff} & 3 & 3 \\
\text{Usually Cooperates} & 2 & \infty \\
\text{Usually Defects} & 2 & \infty \\
\text{Solution B1} & 3 & 2 \\
\text{Solution B5} & 6 & \infty \\
\text{Thumper} & 2 & \infty \\
\text{Evolved FSM 4} & 4 & \infty \\
\text{Evolved FSM 16} & 16 & \infty \\
\text{Evolved FSM 16 Noise 05} & 16 & \infty \\
\text{TF1} & 16 & \infty \\
\text{TF2} & 16 & \infty \\
\text{TF3} & 16 & \infty
\end{array}
\]
\end{table}

\section{Discussion}

We have defined an efficient algorithm for computing the memory-depth of finite state machine strategies that can be
used to understand the complexity of such strategies on equal footing with strategies otherwise implemented or represented.
We applied the algorithm to a set of finite state machine strategies in the Axelrod library,
which includes over 200 strategies and all of the 63 strategies in Axelrod's second tournament
as of library version 4.6.0., some of which were evolved for various environments (see table \ref{memory_lengths}).
As most of these finite state machine strategies are infinite, and these strategies are among the best performers in
the Axelrod library,
our algorithm demonstrates that longer memory strategies can in fact outperform simpler strategies such
as TFT (also included in the library). There may, however, be similarly high performing strategies of smaller memory-depth.
Evolutionary processes can easily change a finite memory strategy into a longer or infinite one.
Given the algorithm presented in this paper, future strategy development could restrict the space of FSM strategies to
those of a maximum memory-depth.

\bibliographystyle{myamsplain}
\bibliography{fsm-memory}

\end{document}